%% file: DHH_RealTime2020.tex
\documentclass{IEEEtran}
\usepackage{cite}
\usepackage{amsmath,amssymb,amsfonts}
\usepackage{graphicx}
\usepackage{import}
\usepackage{textcomp,nicefrac}
\usepackage{xcolor}
\usepackage{siunitx}
\usepackage[bookmarks=false]{hyperref}
\DeclareSIUnit{\Bit}{Bit}
\DeclareSIUnit{\Byte}{Byte}
\DeclareSIUnit{\word}{word}

\newcommand{\belle}{Belle~II}
\newcommand{\secref}[1]{Sec.~\ref{#1}} 
\newcommand{\figref}[1]{Fig.~\ref{#1}} 
\def\BibTeX{{\rm B\kern-.05em{\sc i\kern-.025em b}\kern-.08em
T\kern-.1667em\lower.7ex\hbox{E}\kern-.125emX}}

\usepackage[switch]{lineno}
\begin{document}

\title{Performance of the {\color{black}Data Handling Hub} readout system for the \belle{} pixel detector}
\author{Stefan Huber, Igor Konorov, Dmytro Levit, Stephan Paul, and Dominik Steffen
\thanks{Manuscript submited for review October 30, 2020, revised April 12, 2021}
\thanks{This work was supported by the BMBF, the Excellence Cluster ORIGINS which is funded by the Deutsche Forschungsgemeinschaft (DFG, German Research
Foundation) under Germany’s Excellence Strategy EXC-2094 390783311, the Maier-Leibnitz-Laboratorium der Universit\"at und der Technischen Universit\"at M\"unchen, and Horizon 2020 Marie Sk\l{}odowska-Curie RISE project JENNIFER2 grant agreement No. 822070 (European grants).}
\thanks{S.~Huber, I.~Konorov, D.~Levit, S.~Paul, and D.~Steffen  are with the Technischen Universit\"at M\"unchen, Garching, 85748, Germany (e-mail: stefan.huber@tum.de)}
}
\maketitle

\begin{abstract}

The SuperKEKB accelerator in Tsukuba, Japan is providing $e^+e^-$ beams for the \belle{} experiment since {\color{black}March 2019}. To deal with the aimed peak luminosity being forty times higher than the one recorded at Belle, a pixel detector based on DEPFET technology has been installed. It features a long integration time of \SI{20}{\micro\second} resulting in an expected data rate of \SI{20}{\giga\Byte\per\second} {\color{black}(\SI{160}{\giga\Bit\per\second})}  at a maximum occupancy of \SI{3}{\percent}. To deal with this high amount of data, the data handling hub (DHH) has been developed{\color{black}. I}t contains all necessary functionality for the control and readout of the detector. In this paper we describe the architecture and features of the DHH system. Further we will show the key performance characteristics after one year of operation.
\end{abstract}


\section{Introduction}
\label{sec:introduction}
\import{}{introduction}

\section{The \belle{} pixel detector}
The \belle{} pixel detector\cite{DEPFET} is the innermost structure in the \belle{} experiment. It consists of two layers of in total 40 modules, called half ladders. Two half ladders are glued together and form one ladder. Those are arranged in a cylindrical structure around the interaction point. Currently, 20 modules are installed in the running experiment, this corresponds to the full inner layer of eight ladders and two ladders of the outer layer.

\begin{figure}[t!]
\centerline{\includegraphics[width=3.5in]{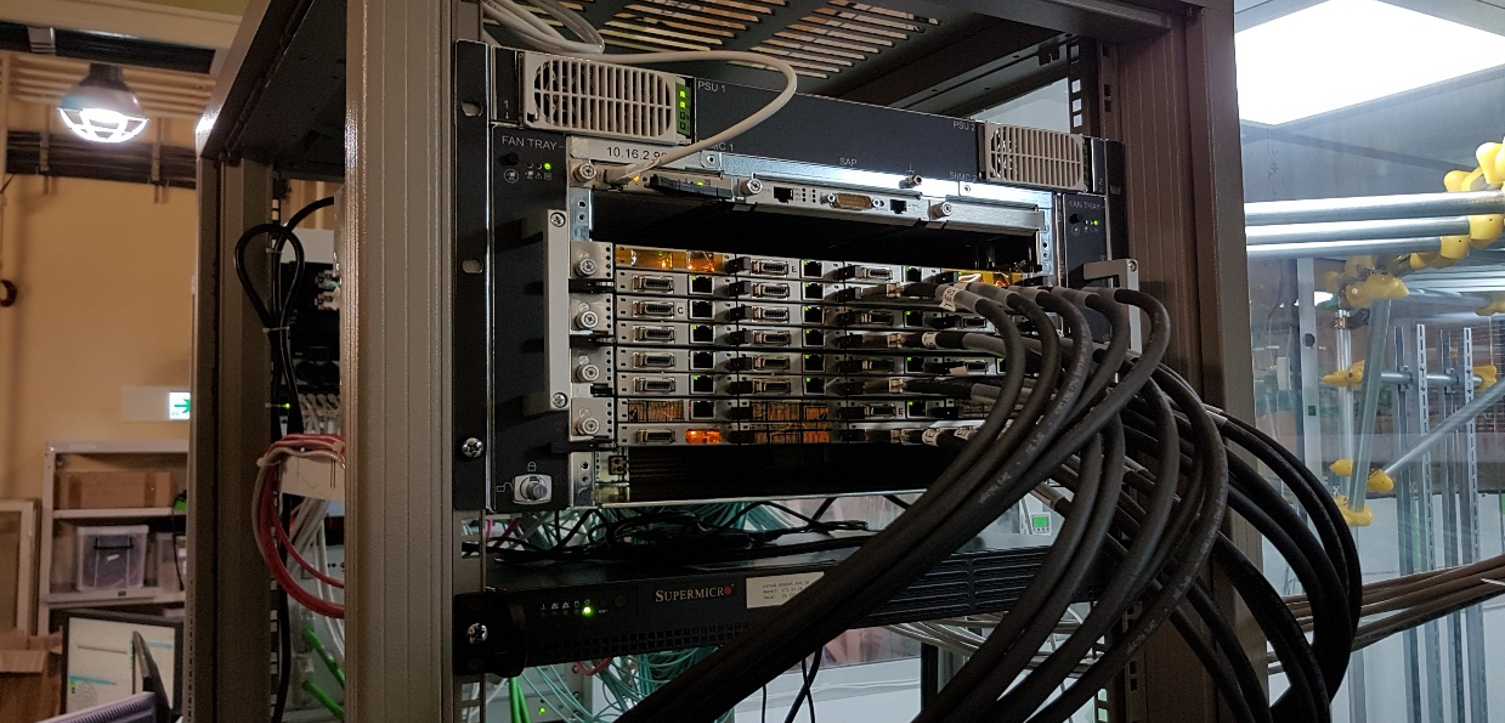}}
\centerline{}
\centerline{\includegraphics[width=3.5in]{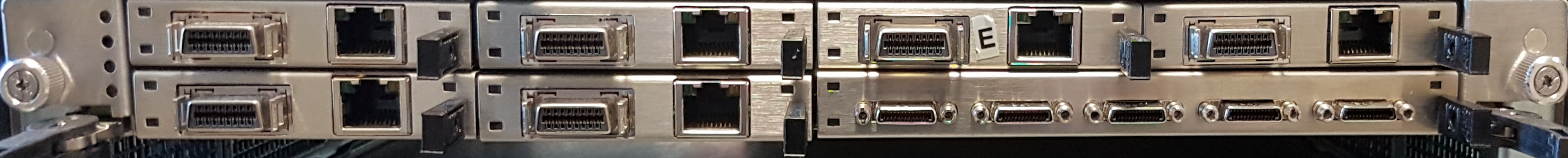}}
\caption{Front view of the DHH system. In this ATCA shelf (top), four DHH carrier cards are installed and connected with CameraLink cables, attached to the DHIs, to the detector. The image on bottom shows a single DHH system. Only the connectors of the DHI (lower right) are used.}
\label{fig:fulldhh}
\end{figure}

The half ladders are built using DEPFET technology and provide a pixel matrix of 768$\times$250 pixels, each. The module is controlled by the SwitcherB \cite{DCD} ASIC. This chip is responsible for the running shutter operation mode, which controls the readout and clearing of the pixel information. The operation of the sensor is performed in groups of four rows, the so-called gates, which are digitized in the same readout cycle.

The readout of the pixel matrix is performed using four pairs of ASICs. Each pair reads one quarter of the columns and consists of the drain current digitizer (DCD) \cite{DCD} and the data handling processor DHPT \cite{DHP}. The DCD digitizes 256 detector signals at once using ADCs with \SI{8}{\Bit} resolution {\color{black}at a sampling rate of \SI{9.5}{\mega\hertz}} and sends the data to the DHPT. On the DHPT the data are processed with algorithms {\color{black}such as} pedestal subtraction, zero suppression, and common mode correction. The data are then further arranged in {\color{black} 8b10b encoded} Aurora{\color{black}\cite{aurora}} frames and sent out to the DHH system, discussed in the rest of the paper. It has to be noted that the limit for lossless data processing on the DHPT is given by a maximum occupancy of \SI{3}{\percent}.

After processing in the DHH{\color{black}, the data are then sent to the online selection nodes (ONSEN) \cite{ONSEN} where} they are buffered until the high-level trigger (HLT) decides to store or discard that event. Further, the HLT extrapolates the tracks measured by the rest of the \belle{} experiment to the surface of the PXD. Only hits close to the extrapolated tracks are sent further to the storage.

 \begin{figure}[t]
\centerline{\includegraphics[width=3.in]{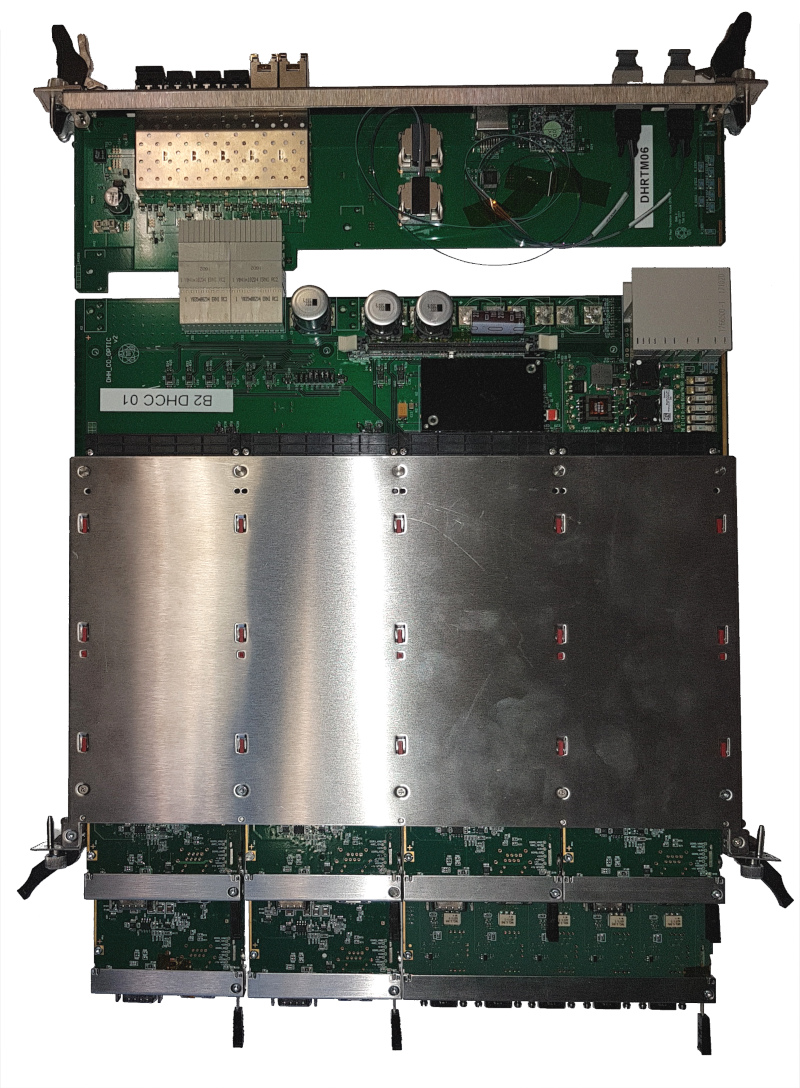}}
\caption{Top view of the DHH with attached RTM. The processing cards are slightly pulled out in order to show the structure.}
\label{fig:dhh_top}
\end{figure}

\section{Hardware}
\label{sec:hardware}
\begin{figure}[t]
\centerline{\includegraphics[width=3.5in]{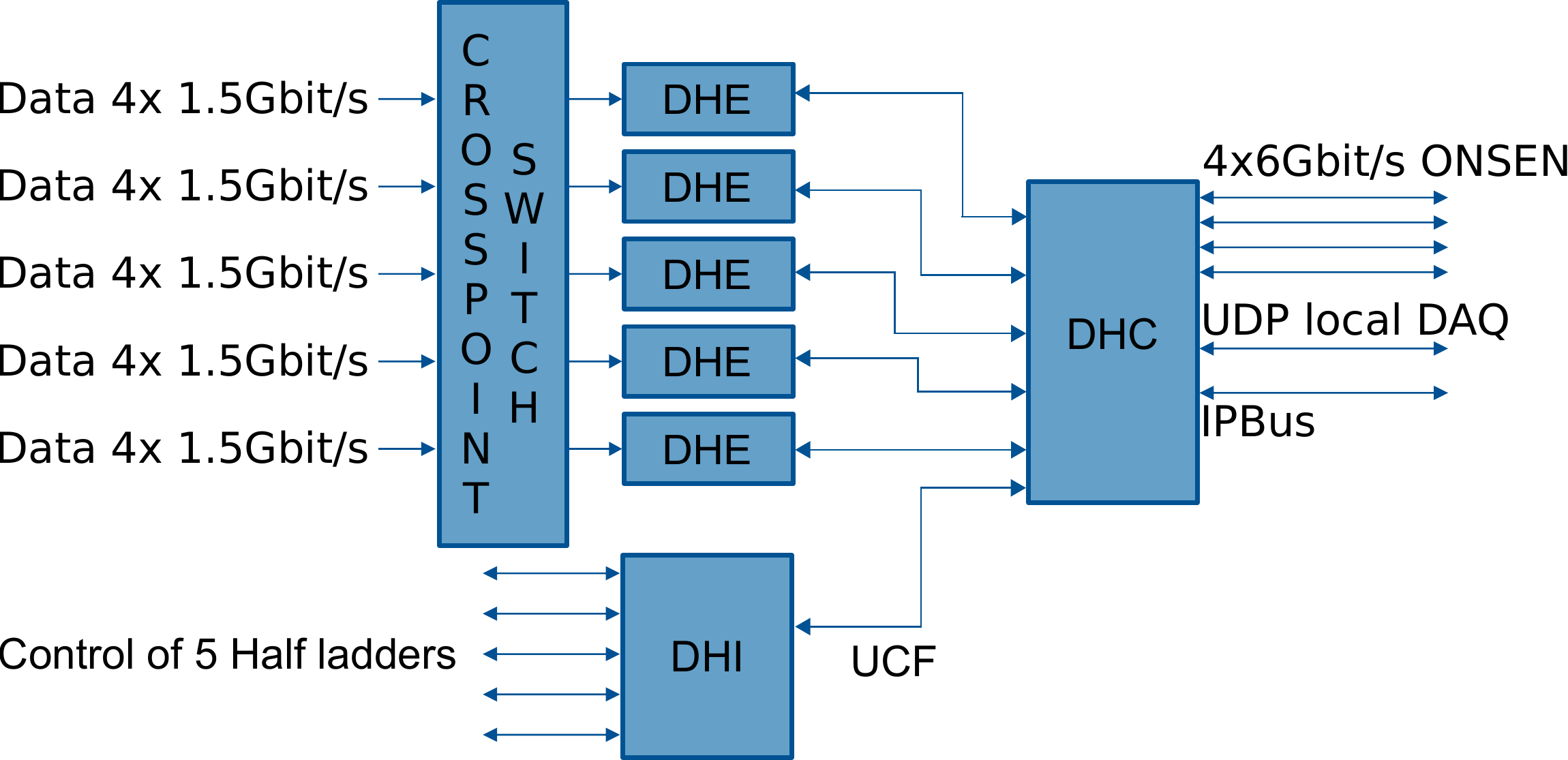}}
\caption{Data flow in the DHH system. {\color{black} The crosspoint switch connects the detector modules to the DHEs. From there they are sent to the attached DHCs and then further to the ONSEN system or local DAQ. Real-time signals and slow-control information is distributed by the DHC and the detector modules are controlled by the DHI, using that data.}}
\label{fig:dhhflow}
\end{figure}

The DHH system, {\color{black}shown in \figref{fig:fulldhh} and  \figref{fig:dhh_top},} is built based on the ATCA standard\cite{atca} and utilizes eight ATCA carrier-cards for the readout of the full PXD.
We equip each of these carriers with seven AMC cards: one DHC, five DHEs, and one DHI. We designed two of them, the data handling engines (DHE) and the data handling concentrator (DHC), as single-width AMC modules. {\color{black} The data handling insulator (DHI) has been designed in the double-width form-factor by us}. The interconnection in the system is set-up in a star topology where the DHC serves as the master and is connected to each of the other modules with a single high-speed serial link. The data flow is depicted in \figref{fig:dhhflow}.

External interfaces are implemented using rear transition modules (RTMs), they are equipped with six SFP+ transceivers. Four of them are used for sending out the data to the DAQ, one for transferring monitoring data to a local DAQ, and one for slow-control. The detector data are received by two MiniPOD receivers which are capable of handling 20 data interfaces to the detectors. Control signals are distributed to the detectors using five 20 meters long CameraLink{\color{black}\cite{cameralink}} cables attached to the front panel of the DHI.
\begin{figure}[t]
\centerline{\includegraphics[width=3.5in]{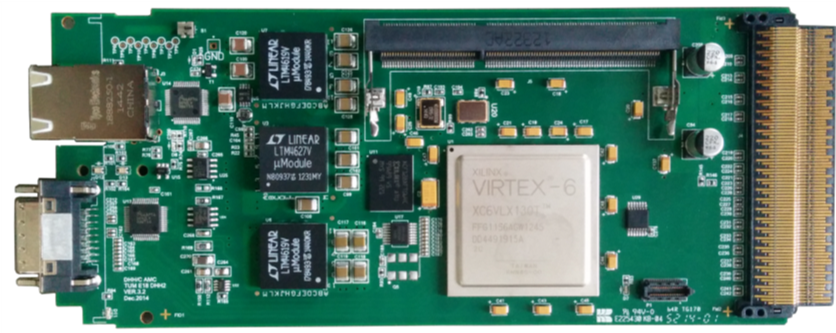}}
\caption{Layout of the AMC module used for the DHE and DHC cards. }
\label{fig:DHEAMC}
\end{figure}
The DHE and DHC modules are based on the same hardware design \figref{fig:DHEAMC}, their functionality is implemented using a Virtex-6 VLX130T FPGA.
Via the AMC connector, a maximum data rate of \SI{5}{\giga\Bit\per\second} can be achieved between the modules while a maximium data rate of \SI{6.5}{\giga\Bit\per\second} is achieved via the RTM.
Both cards are equipped with \SI{4}{\giga\Byte} of DDR3 memory mounted on a SO-DIMM slot. A configurable clock generator, Si5338, is used to reduce the jitter of the recovered clock from the \belle{} time-distribution system as well as for the clocks recovered from the high-speed links {\color{black} to \SI{14}{\pico\second}}. 

\import{}{DHE}

\import{}{DHC}
\import{}{DHI}

\subsection{The Crosspoint Switch}
The PXD read out system is equipped with a 144$\times$144 crosspoint switch, originally developed for the DAQ of the COMPASS experiment \cite{Steffen:2018tdp}. The switch provides full flexibility to define optical links' topology between the detectors and the DHH. The ONSEN system needs a defined order of the detector information in the data stream which matches the order in which the regions of interest are received from the HLT. This does not coincide with the topology in which the detector is cabled. Installation of the switch gives several other advantages over a direct connection such as changeover in case of failure or mirroring the links on another DHE for monitoring or testing new features. The switch allows to monitor the optical power of each link with high precision providing good diagnostics of the optical transmitters installed close to the detector and exposed to ionizing radiation.

\subsection{The UCF Protocol}
\begin{figure}[h]
\centerline{\includegraphics[width=3.5in]{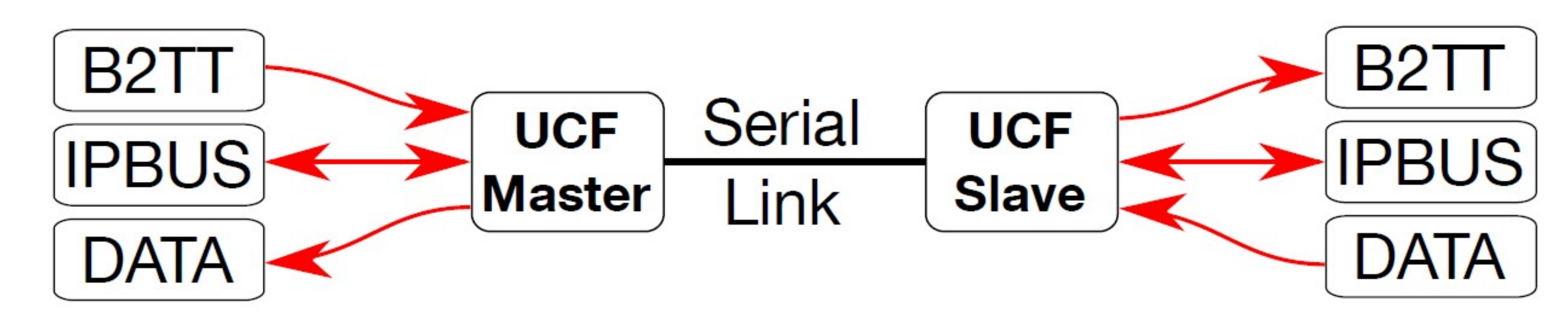}}
\caption{Data stream connected to the UCF link. B2TT represents the real-time signals transmitted to the DHE/DHI and includes information like triggers, synchronization, and resets.}
\label{fig:ucf}
\end{figure}

For the data links between DHC and DHI/DHE we utilize the unified communication framework UCF\cite{GaisbauerUCF}. This is a protocol developed for particle physics  \cite{Gaisbauer:2016txp} experiments. {\color{black}It provides a clock with a jitter, measured to be \SI{23}{\pico\second} in our system }at the receiving side together with a deterministic phase relation. It is possible to transmit up to 64 different data streams via a single serial link. One stream has a deterministic latency and shall be used to transmit real-time signals of the time distribution system. In our application, we use two additional channels, one in a bidirectional way to transmit UDP frames used for the slow-control by IPbus \cite{Frazier:2012iva} and one for transmission of the data, \figref{fig:ucf}. {\color{black}In order to catch transmission errors the data protocol has a CRC32 checksum attached to every frame. Errors in operation are prevented by attaching  a CRC16 checksum to every command word of the real-time channel, in case of a checksum mismatch the command is ignored. In that case, missing events are recovered on the DHC.}

\subsection{Synchronization Of The System}
\label{sec:sync}
All detectors in the \belle{} experiment are synchronized with the SuperKEKB accelerator by \belle{} Trigger and Time distribution system (B2TT) \cite{Nakao:2012zz}. The B2TT system broadcasts a \SI{127.21}{\mega\hertz}
clock that is generated from the accelerator’s radio frequency clock of \SI{508.84}{\mega\hertz}. The trigger and synchronization information is broadcasted together with the B2TT clock using an 8b/10b encoding over serial lines.

The DHC as well as the interfaces to DHE and DHI are using the recovered \SI{127.21}{\mega\hertz} clock as the main time reference. The detector is operating at a lower frequency of \SI{76.33}{\mega\hertz} derived from the main clock by fractional divider 3/5. This clock is synthesized locally on the corresponding modules and special care is taken to transfer all real-time signals synchronously between the two clock domains. It is important to note that the control signals to the detector are sent-out only every 8th clock cycle of \SI{76.33}{\mega\hertz} and thus the signals have to be synchronized between the two clock domains in this time frame. Taking this into consideration the transferred signals are sent out from the DHC in three fixed time intervals of 13, 13 and 14 clock cycles in the \SI{127.21}{\mega\hertz} domain. This sequence repeats every 40 clock cycles and requires a minimum distance between two triggers of about \SI{120}{\nano\second}.

\import{}{performance}

\section{Conclusion and Outlook}
During the last two years, we have proven that we can reliably read out the \belle{} pixel detector with the DHH system. The next major milestones are the installation of the full PXD in 2022 which requires the installation of four additional DHH modules at \belle{}. The trigger rates will rise in the future and thus the link speeds on UCF have to be increased to \SI{5}{\giga\Bit\per\second} to cope with the increased data rate. We are currently redesigning the carrier card to perform this mandatory step.

An important next step, not directly related to the DHH, is the commissioning of the gated mode. Here, work is ongoing in terms of analysis and detector optimization.

Two additional upgrades for the DHH are planned. Currently we use UDP for reading out data to the local DAQ. As UDP does not have any error handling, part of the data is lost on that interface. We are currently investigating a changeover to a readout based on a commercial PCIe card which will allow us to reliably record these data.

Finally, we plan to implement a clustering algorithm in the DHE. Based on the parameters of these clusters it is possible to keep hits in the data-stream which otherwise would be disregarded by the HLT decision \cite{Baehr:2015tbf}. This is especially important for hits from slow pions which will not reach the outer tracking detector of the \belle{} experiment.

\bibliographystyle{IEEEtran}
\bibliography{IEEEabrv,DHH_RealTime2020}


\end{document}

%% file: introduction.tex
\IEEEPARstart{S}{ince} the beginning of 2019 the \belle{} experiment is taking data. 
The experiment is studying events from electron-positron collisions with beam energies of \SI{7}{\giga\electronvolt} for electrons and \SI{4}{\giga\electronvolt} for positrons. The design luminosity is 40 times higher than the one at the Belle experiment and shall reach up to \SI{8e35}{\per\centi\meter\squared\per\second}. The experiment is built in a shell structure around the interaction point and consists of three types of tracking detectors, namely{\color{black},} the pixel detector (PXD), the silicon vertex detector (SVD), and the central drift chamber (CDC). These are surrounded by detectors for particle identification and calorimetry.

In this paper we describe the implementation and performance of the readout system of the PXD, the data handling hub (DHH).
It is the most demanding part in the readout chain of the \belle{} experiment as it has to deal with \SI{95}{\percent} of the data.

%% file: DHE.tex
\subsection{The data  handling  engine}
\begin{figure}[t]
\centerline{\includegraphics[width=3.5in]{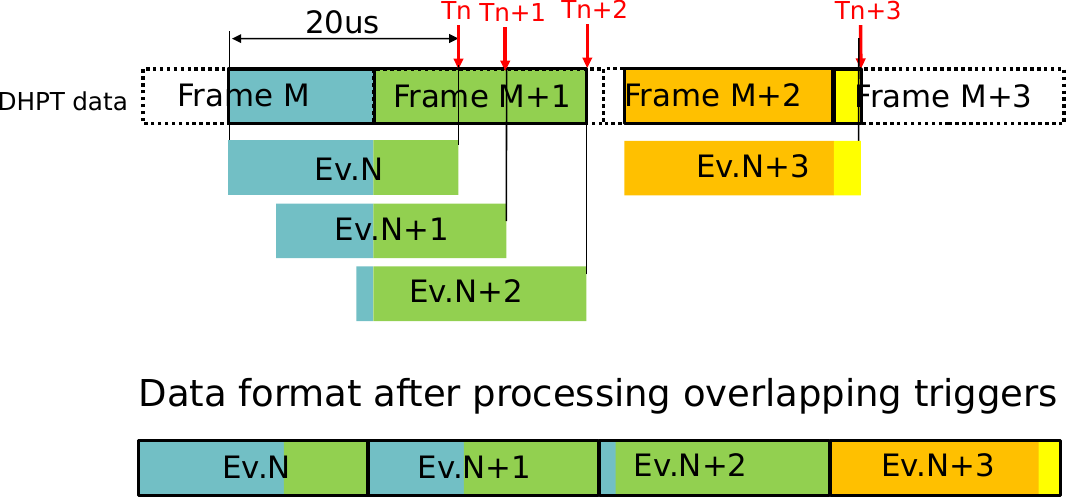}}
\caption{Division of the data in case part of the data belongs to several different triggers.}
\label{fig:overlapping}
\end{figure}
\begin{figure}[t]
\centerline{\includegraphics[width=3.5in]{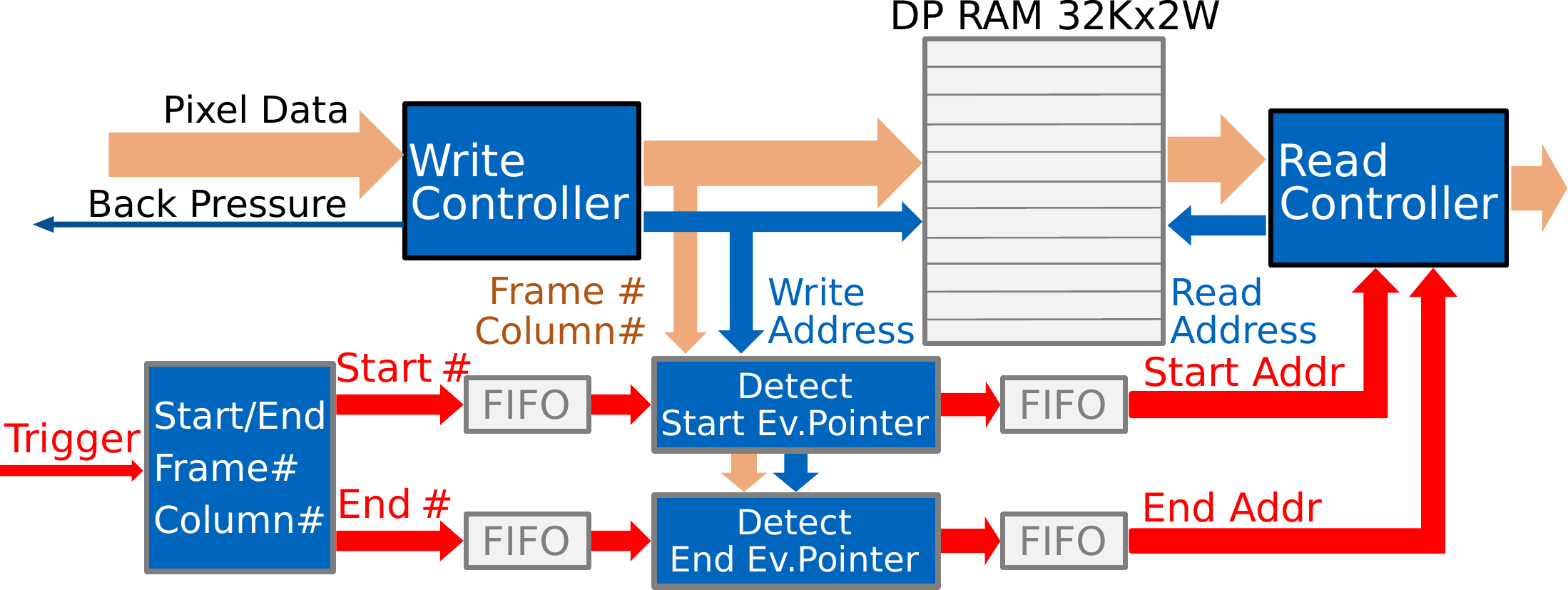}}
\caption{Data processing scheme on the DHE.{\color{black}, which converts streamed data to event-based data format.}}
\label{fig:overl_flow}
\end{figure}
The purpose of the DHE is to process the data from one half-ladder of the PXD. The data are received via four optical links running with a \SI{1.5}{\giga\Bit\per\second} Aurora {\color{black}(8b10b)} link, each. These data are pre-buffered using \SI{4}{\giga\Byte} of DDR3-memory organized as four FIFOs, one for each incoming data stream. 

The challenge on this module comes from the long integration time of \SI{20}{\micro\second} of the detector together with trigger rates up to \SI{30}{\kilo\hertz}. The DHPT ASIC has a simple logic to handle these requirements. The detector information is processed by DHPT continuously with the speed of the running shutter. The trigger signal opens a \SI{20}{\micro\second} gate enabling the DHPT to send data out. If the following trigger is closer than \SI{20}{\micro\second} then the gate signal is extended for the next \SI{20}{\micro\second}. A task of the DHE is to synchronize DHPT data with a trigger by using frame and row numbers information transmitted by the DHPT.
When the time between triggers is smaller than \SI{20}{\micro\second}, a part of detector data shall be shared between two or more consecutive events, \figref{fig:overlapping}. This is resolved by analyzing the time of arrival of the individual triggers. The DHPT emulator, implemented in the DHE, predicts which part of the readout frames corresponds to which triggeri, as shown in \figref{fig:overl_flow}. Using this information the data are analyzed by corresponding state machines and start/end address pointers of each event are detected. Then they are written to the dual-port random-access memory. Subsequently the pointers are transmitted to the read controller for event data extraction from this memory. There are four of such data processing streams in the DHE, one for each DHPT ASIC. At the last step of the data processing the data from four streams are merged in one sub-event and sent to the DHC. 

The data processing in the DHE is capable to store up to 2000 events in the pipeline. The DHC issues a busy signal to the \belle{} trigger system to prevent buffer overflow and loss of data.

In contrast to the DHE{\color{black},} the DHPT ASICs do not have a capability to throttle triggers in case {\color{black}the} data rate exceeds the bandwidth of outgoing interfaces. This may happen  in the first few revolution cycles after bunch injection, due to high background. In that situation the data may be scrambled and can affect the DHE data processing algorithm. The DHE performs a sanity data check at the input to minimize an impact of such data on stability of data taking. Although not all data with errors are excluded from the readout, full error diagnostic is provided via slow control.

%% file: DHC.tex
\subsection{The  data  handling  concentrator}
\begin{figure}[t]
\centerline{\includegraphics[width=3.5in]{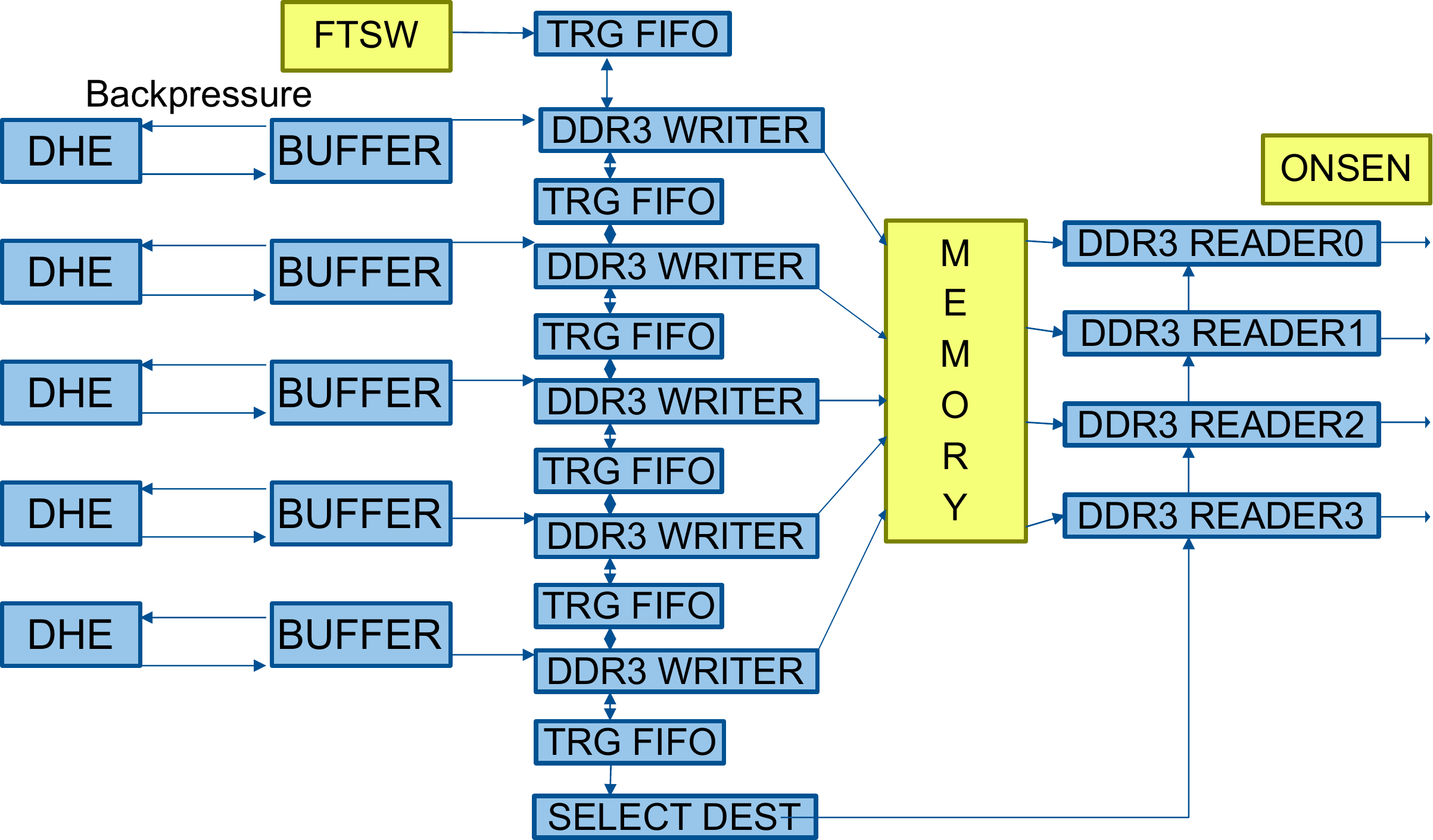}}
\caption{Data processing scheme on the DHC.}
\label{fig:dhcflow}
\end{figure}

The data processing on DHC has two different duties. As a first task, it has to provide a 5-to-1 multiplexer to merge data from 5 DHE modules and, as a second task, it has to distribute the events in a round-robin manner between four ONSEN nodes. As these two functionalities are closely coupled, they are implemented in a common framework taking advantage of buffering in the DDR3 memory. The data flow is depicted in \figref{fig:dhcflow}.

In the first stage, the input data are pre-buffered in a local \SI{4}{\kilo\word} deep FIFO which issues back pressure to the DHE in case it fills up above \SI{95}{\percent}. The data are written into the memory in a sorted way. This means the data from the first DHE for one event are written, then the data for the second one till finally the data from the last DHE are stored in the memory. The pointers from the different DHEs are passed from one writer state machine to another so after one cycle  a fully assembled event is stored in the memory. Each event is stored in a predefined memory slot which can hold up to \SI{1}{\mega\Byte} of data. Due to this mechanism, the DHEs with a higher index have to wait until the DHEs with a lower index have finished processing the current event. This creates the necessity to buffer the events on the DHEs to avoid any loss of data. 

All five inputs are loosely synchronized between each other by transmitting pointers via trigger fifos. It allows to compensate possible interference between streams due to variation of data rates and achieve fully parallel data processing very close to the maximum speed.

The memory-pointers of the events are then passed to an output-selection state machine which decides to which ONSEN node the event shall be written. This is done based on a lookup table checking the four least-significant bits of the event number and which ONSEN nodes are currently used in the system.

Finally, one out of four data-reader state machines receives the pointers and reads a full event from the memory. Together with the trigger information a full event is assembled and sent out via a \SI{6.5}{\giga\Bit\per\second} Aurora link to the ONSEN system.

Part of the data is needed for real-time monitoring of the system performance. To achieve that, the first output of the event builder is connected to two interfaces. One of them takes only a small fraction of the data and sends them via a UDP link to the local DAQ. In the case of detector studies, it is possible to send data only via this UDP link. In this mode of operation the maximum trigger rate is limited by the UDP bandwidth.

The DHC is designed in a way that it can handle up to 1000 triggers in a pipeline. In case a threshold of 900 triggers is exceeded it sends a 'busy' signal to the trigger system. While this signal is active the trigger logic does not generate new triggers.

%% file: DHI.tex
\subsection{The The data handling insulator }
\begin{figure}[t]
\centerline{\includegraphics[width=3.5in]{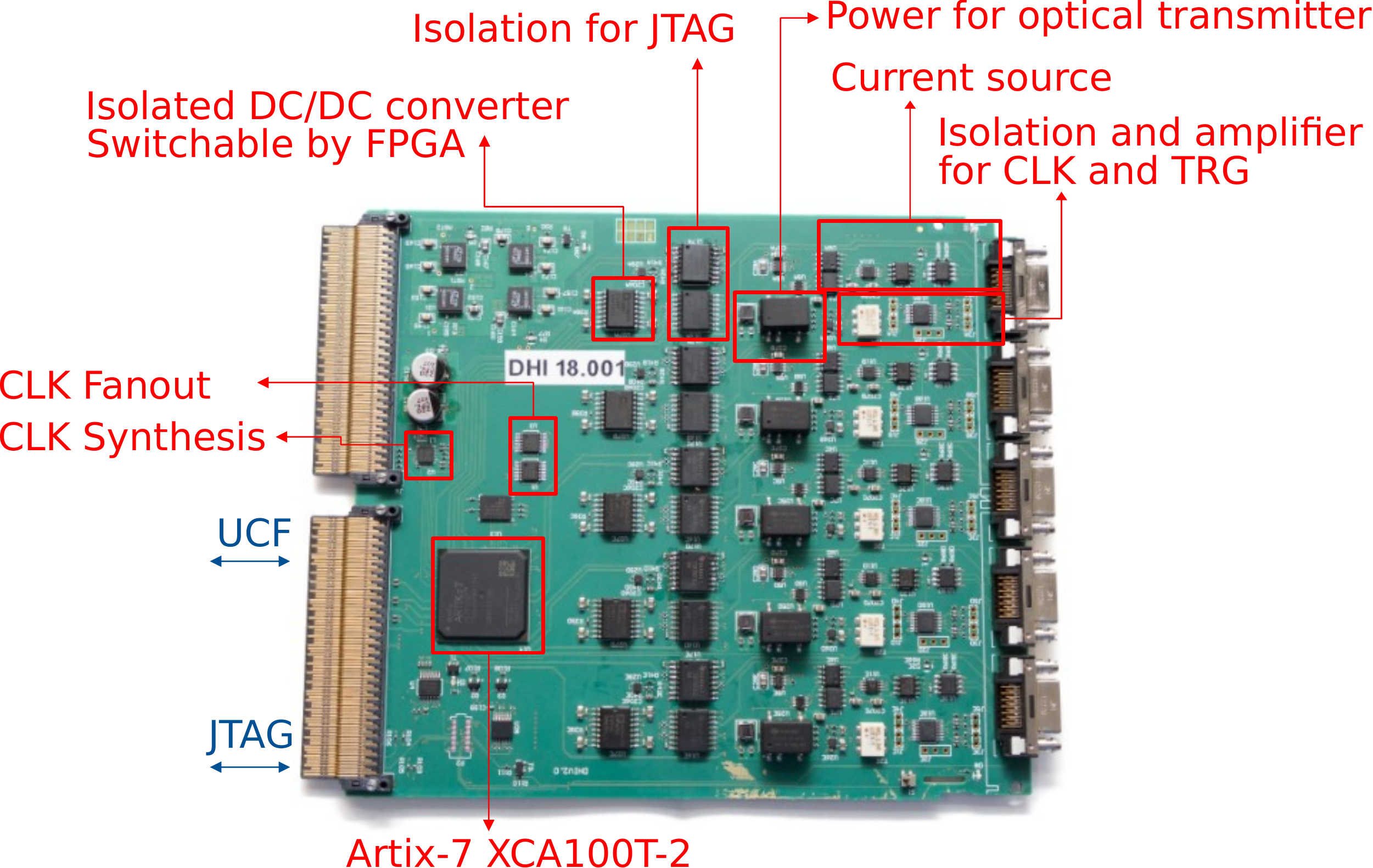}}

\caption{Picture of the DHI module with all important components. Labeled are the components needed to supply one single half ladder.}
\label{fig:DHIAMC}
\end{figure}

The DHI, \figref{fig:DHIAMC}, is the control module of five half-ladders. It distributes the clock signal, control commands, and JTAG slow-control information via the attached CameraLink cable.

The control information is transmitted to the detector using a modified Manchester code with a length of eight clock cycles: this corresponds to four-bit of payload information.
Each of the four bits has an individual function: reset, veto, trigger, and synchronization.
During idle operation, it sends the binary code '00011101', which is used for a control word alignment by the receiving sides. 
 Besides, a special command, which initiates the read out of one full detector frame without zero-suppression, is encoded as  a broken code of the first six bits set to '111000' followed by the encoded synchronization signal.

The reset bit represents two different types of reset distinguished by the length of the active reset sequence, thus it's either called short or long reset. 
The short reset is encoded as active reset bit in only one consecutive control word, it is the weaker of the two types of resets and affects only the high-speed links towards the DHE.
If the reset bit is active for longer than one control word, the long reset is recognized. This reset acts on all logic of the DHPT, especially the gate and frame counters, as well as the high-speed links.
Both resets may be issued via a slow-control command. {\color{black}The long reset is also executed when a start-of-run reset from the Belle 2 Trigger and Time distribution system (B2TT) is received,} see \secref{sec:sync}\cite{Nakao:2012zz}. It is needed to synchronize the DHPTs with the DHPT emulator in the DHE.

The veto bit is used for switching the detectors to the `gated mode'. This functionality is described in more detail in \secref{sec:gatedmode}.

The trigger bit is usually active for \SI{20}{\micro\second} after receiving the B2TT trigger. It corresponds to one full detector frame but can be shorter or longer, in special cases. Activating trigger signal for one command corresponds to about \SI{104}{\nano\second} and enables 4 detector rows to be read out.

The synchronization bit synchronizes the running shutter to the revolution cycles of the accelerator. As one PXD frame corresponds to two revolution cycles, the B2TT provides a designated signal synchronously to every other revolution cycle.

The DHI allows us to configure half-ladders over JTAG protocol using 5 JTAG player cores in FPGA. The JTAG players are controlled by an input-output control software (IOC) running on the remote PC.
The JTAG player core communicates with the IOC over a dedicated IPbus instance in the DHI, decodes the bitstream, shifts it into the JTAG chain, records the read back data from the detector, and sends them to the control software.
The IOC communicates with the detector slow-control via the Experimental Physics and Industrial Control System (EPICS) \cite{EPICS}.
{\color{black} EPICS is a framework which provides a register-based network protocol used to design distributed slow control systems. 
The IOC translates the high-level EPICS transaction into low-level commands for the JTAG player.}
The IOC memorizes the content of the JTAG registers and bit fields, which are available for read and write access over EPICS. 
The IOC uses these values to assemble the bitstream when the JTAG access is scheduled, and sends the bitstream to FPGA for execution.
After the transaction is finished, the IOC decodes the detector reply, stores the new values, and makes them available over EPICS.

In addition to the control features, the DHI provides a precise current source, it can be used for calibration of the detector. Further, the DHI supplies the optical transmitter close to the detector with the voltage \SI{3.3}{\volt}. It is possible to switch off all channels via the FPGA in order to avoid powering of the {\color{black} DHPT ASIC by DHI via the LVDS JTAG wires which provide a constant common mode voltage.}

%% file: performance.tex
\section{performance}
\subsection{Gated Mode}
\label{sec:gatedmode}

\begin{figure}[t]
\centerline{\includegraphics[width=3.5in]{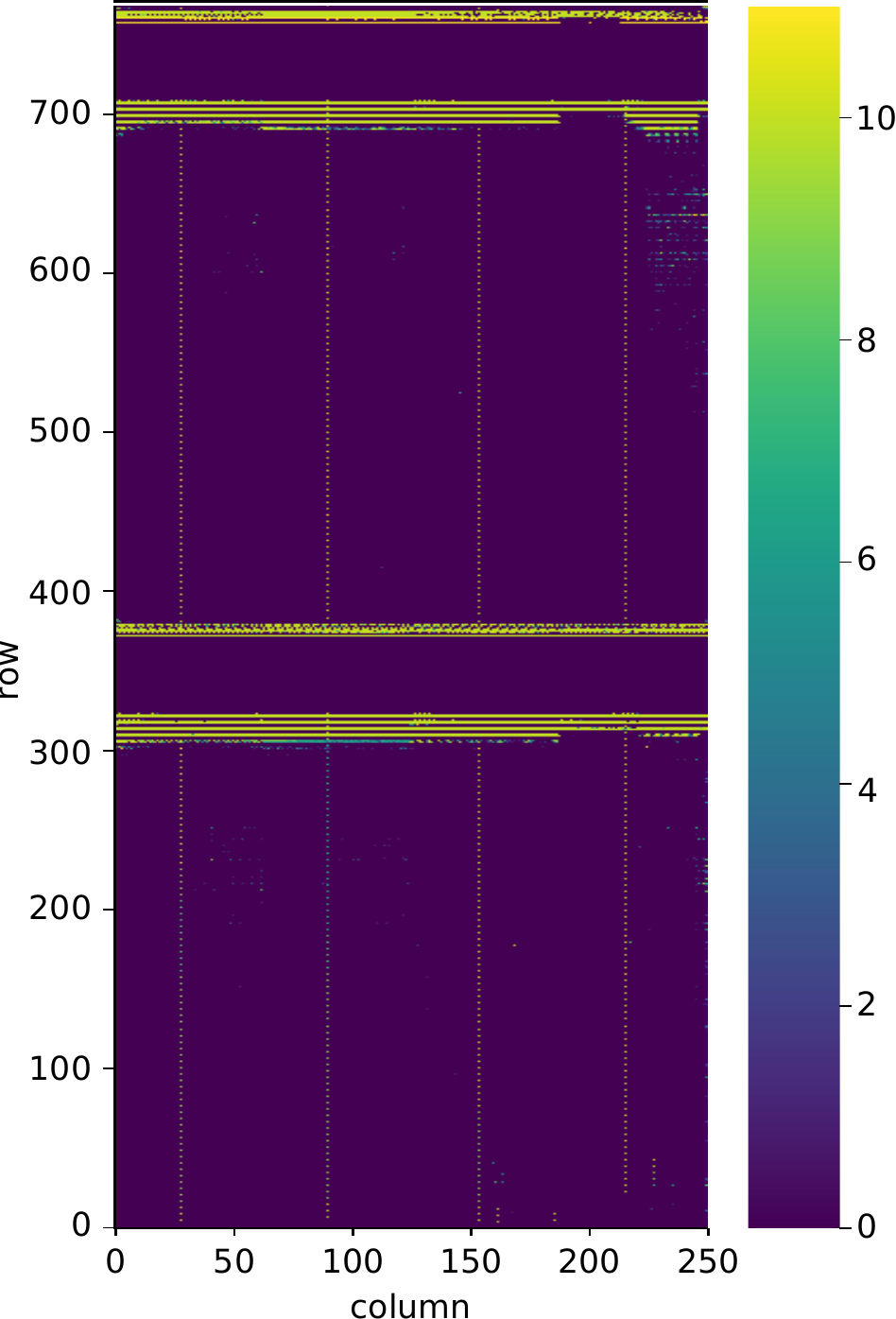}}
\caption{Distribution of hits during a lab measurement with activated gated mode with ten recorded events. There are two regions in which the gated mode was switched on which show an increased occupancy, indicated by the yellow color. Inside these regions bands with now hits due to the deactivated trigger signal are visible. The dotted vertical lines are pixels where zero-supression is switched off, they are used as an optical guide. {\color{black} Most of the additional hits in the matrix are due to the induced pedestal  oscillation as well as from hot pixels which are not taken care of in the lab setup.}}
\label{fig:gatemode}
\end{figure}

The lifetime of the beam in SuperKEKB is approximately \SI{600}{\second} \cite{Bona:2007qt}. In order to operate the experiment at high luminosity, a continuous injection scheme is applied. For each ring bunches of electrons or positrons are injected with a frequency of \SI{25}{\hertz}. These freshly injected bunches lead to an increased background level for a few \SI{}{\milli\second} after their injection.

To make the PXD insensitive to this injection noise, the detector {\color{black}will be} operated in gated mode\cite{gatedmode}.
While in gated mode, the detector stores the already accumulated charge in each pixel but does not
store charge from radiation impinging on the detector. 

As soon as a new bunch is injected into the accelerator the gated mode logic on the DHC is armed by the kick signal of the injection magnet, distributed by the B2TT.
The DHC is responsible for the timing of the gated mode operation, and sends a start as well as a stop signal to the DHI.
The start signal is issued shortly after the kick signal and the stop signal is issued as soon as the injection background is assumed to be low enough.
Typically this is the case after a few~\SI{}{\milli\second}. Both parameters may be tuned individually for both detector rings.

The DHI uses these signals to control the gated mode on the detector by modifying two of the control signals: trigger and veto. The veto bit is activated every revolution cycle (twice per detector frame)
at the time the noisy bunch needs to pass by the detector. An optimal value for this time is the readout duration of 10 gates corresponding to \SI{1.04}{\micro\second}. Gated mode operation 
significantly changes the electrical fields applied to the sensor and thus shifts the pedestal of the pixel which are activated by the rolling shutter close to the time when the gated mode is enabled.
To avoid data loss due to high detector occupancy caused by the corresponding increased noise, the trigger signal is not issued during that period. The typical time constant for this pause is the duration of the readout of 20 gates or \SI{2.08}{\micro\second}. The effect is seen in \figref{fig:gatemode}.

During 2020 operation the gated mode has been tested \cite{Spruck:2020voo} and the principle was proven. Still, a final tuning of the parameters as well as a detailed understanding of the pedestal oscillation has to be performed before it can be activated during data taking.

 \begin{figure}[t]
\centerline{\includegraphics[width=3.in]{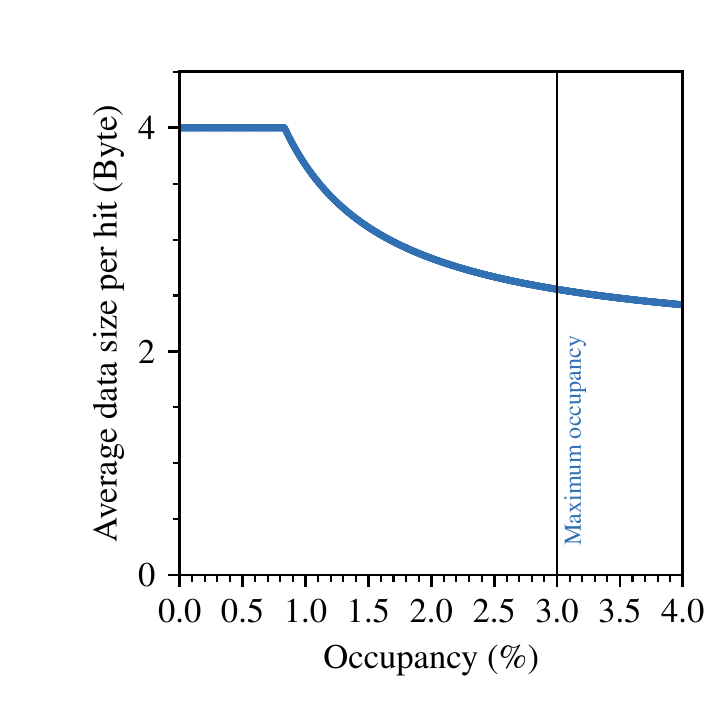}}
\centerline{\includegraphics[width=3.in]{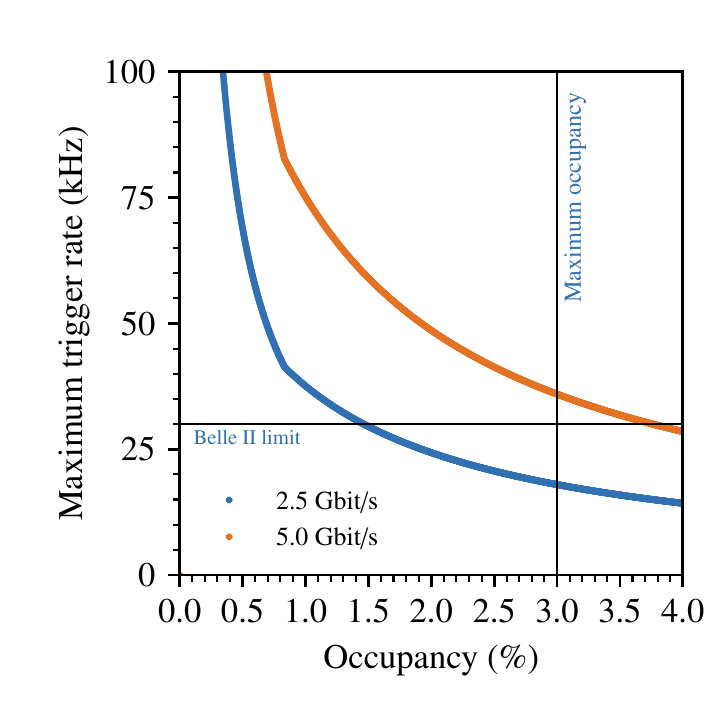}}
\caption{{\color{black}Simulation of the average number of bytes per hit (top)} and maximum trigger rate {\color{black}(bottom)}  as a function of the detector occupancy. In this calculation the most inefficient distribution of hits in terms of the data format is assumed. The maximum occupancy is due to the limits of the DHP while the \belle{} limit is the anticipated trigger rate. {\color{black}The orange line is the calculated trigger rate limit in case a \SI{5.0}{\giga\Bit\per\second} is used.} The limitation of the trigger rate is due to the links speed between DHE and DHC.}
\label{fig:perf}
\end{figure}

\subsection{Performance During \belle{} Operation}

{\color{black} 
We installed and operated four DHH carriers at the \belle{} experiment and operated them successfully during the runs in 2019 and 2020.
As we are still in a phase where the performance of the accelerator is optimized, the trigger rates of \SI{5}{\kilo\hertz} are still significantly below the design rate of \SI{30}{\kilo\hertz} which is anticipated for the future. Further, the average detector occupancy was below \SI{0.5}{\percent} due to the not yet final luminosity of the accelerator.

In these conditions the average data rate for a single DHE was \SI{5}{\mega\Byte\per\second} and \SI{94}{\mega\Byte\per\second} for the installed system consisting of 20 DHEs. Peak data rates were \SI{20}{\mega\Byte\per\second} for the DHE with the highest occupancy and \SI{160}{\mega\Byte\per\second} for the current system.

During test runs with artificial triggers and generated data on the DHPT we reached a maximum throughput of \SI{250}{\mega\Byte\per\second} per DHE and in total \SI{5}{\giga\Byte\per\second}. This maximum throughput limits the system performance to \SI{17}{\kilo\hertz} at a detector occupancy of \SI{3}{\percent}. This corresponds to a throughput of \SI{10}{\giga\Byte\per\second} for the final system.

The throughput of the DHH is significantly above the requirements of the current operation but still a factor two below the designed performance.
In order to meet the future requirements a redesign of the ATCA carrier is ongoing. 

}

%% file: DHH_RealTime2020.bbl
\begin{thebibliography}{10}
\providecommand{\url}[1]{#1}
\csname url@samestyle\endcsname
\providecommand{\newblock}{\relax}
\providecommand{\bibinfo}[2]{#2}
\providecommand{\BIBentrySTDinterwordspacing}{\spaceskip=0pt\relax}
\providecommand{\BIBentryALTinterwordstretchfactor}{4}
\providecommand{\BIBentryALTinterwordspacing}{\spaceskip=\fontdimen2\font plus
\BIBentryALTinterwordstretchfactor\fontdimen3\font minus
  \fontdimen4\font\relax}
\providecommand{\BIBforeignlanguage}[2]{{%
\expandafter\ifx\csname l@#1\endcsname\relax
\typeout{** WARNING: IEEEtran.bst: No hyphenation pattern has been}%
\typeout{** loaded for the language `#1'. Using the pattern for}%
\typeout{** the default language instead.}%
\else
\language=\csname l@#1\endcsname
\fi
#2}}
\providecommand{\BIBdecl}{\relax}
\BIBdecl

\bibitem{DEPFET}
C.~Marinas, ``{The Belle II pixel detector: High precision with low
  material},'' \emph{Nucl. Instrum. Meth. A}, vol. 731, pp. 31--35, 2013.

\bibitem{DCD}
I.~Peri\'c, P.~Fischer, T.~H.~H. Nguyen, and L.~Knopf, ``{DCDB and SWITCHERB,
  the readout ASICS for BELLE II DEPFET pixel detector},'' in \emph{{2011 IEEE
  Nuclear Science Symposium and Medical Imaging Conference}}, 2011, pp.
  1536--1539.

\bibitem{DHP}
M.~Lemarenko, T.~Hemperek, H.~Kr\"uger, M.~Koch, F.~L\"utticke, C.~Marinas, and
  N.~Wermes, ``{Test results of the data handling processor for the DEPFET
  pixel vertex detector},'' \emph{JINST}, vol.~8, p. C01032, 2013.

\bibitem{aurora}
\BIBentryALTinterwordspacing
Aurora 8b/10b. Xilinx, Inc. [Online]. Available:
  \url{https://www.xilinx.com/products/intellectual-property/aurora8b10b.html}
\BIBentrySTDinterwordspacing

\bibitem{ONSEN}
T.~Ge\ss{}ler, W.~K\"uhn, J.~S. Lange, Z.~Liu, D.~M\"unchow, B.~Spruck, and
  J.~Zhao, ``{The ONSEN Data Reduction System for the Belle II Pixel
  Detector},'' \emph{IEEE Trans. Nucl. Sci.}, vol.~62, no.~3, pp. 1149--1154,
  2015.

\bibitem{atca}
\BIBentryALTinterwordspacing
\emph{AdvancedTCA\textregistered~Overview}, PICMG (PCI Industrial Computer
  Manufacturers Group) Std. [Online]. Available:
  \url{https://www.picmg.org/openstandards/advancedtca/}
\BIBentrySTDinterwordspacing

\bibitem{cameralink}
\BIBentryALTinterwordspacing
\emph{Camera Link – The Only Real-Time Machine Vision Protocol}, AIA
  (Automated Imaging Association) Std. [Online]. Available:
  \url{https://www.visiononline.org/vision-standards-details.cfm?type=6}
\BIBentrySTDinterwordspacing

\bibitem{Nakao:2012zz}
M.~Nakao, ``{Timing distribution for the Belle II data acquistion system},''
  \emph{JINST}, vol.~7, p. C01028, 2012.

\bibitem{EPICS}
L.~R. Dalesio, J.~O. Hill, M.~Kraimer, S.~Lewis, D.~Murray, S.~Hunt, W.~Watson,
  M.~Clausen, and J.~Dalesio, ``The experimental physics and industrial control
  system architecture: past, present, and future,'' \emph{Nuclear Instruments
  and Methods in Physics Research Section A: Accelerators, Spectrometers,
  Detectors and Associated Equipment}, vol. 352, no.~1, pp. 179 -- 184, 1994.

\bibitem{Steffen:2018tdp}
S.~Huber \emph{et~al.}, ``{Intelligence Elements and Performance of the
  FPGA-based DAQ of the COMPASS Experiment},'' \emph{PoS}, vol. TWEPP-17, p.
  127, 2018.

\bibitem{GaisbauerUCF}
D.~Gaisbauer, Y.~Bai, S.~Huber, I.~Konorov, D.~Levit, S.~Paul, and D.~Steffen,
  ``{Unified communication framework},'' in \emph{{20th IEEE-NPSS Real Time
  Conference}}, 2016.

\bibitem{Gaisbauer:2016txp}
D.~Gaisbauer, Y.~Bai, I.~Konorov, S.~Paul, and D.~Steffen, ``{Self-triggering
  readout system for the neutron lifetime experiment PENeLOPE},'' \emph{JINST},
  vol.~11, no.~02, p. C02068, 2016.

\bibitem{Frazier:2012iva}
R.~Frazier, G.~Iles, D.~Newbold, and A.~Rose, ``{Software and firmware for
  controlling CMS trigger and readout hardware via gigabit Ethernet},''
  \emph{Phys. Procedia}, vol.~37, pp. 1892--1899, 2012.

\bibitem{Bona:2007qt}
M.~Bona \emph{et~al.}, ``{SuperB: A High-Luminosity Asymmetric e+ e- Super
  Flavor Factory. Conceptual Design Report},'' 2007.

\bibitem{gatedmode}
M.~Valentan, E.~Prinker, F.~M\"uller, C.~Koffmane, and R.~Richter, ``{Gated
  mode operation of DEPFET sensors for the Belle II pixel detector},'' in
  \emph{{2015 IEEE Nuclear Science Symposium and Medical Imaging Conference}},
  2016, p. 7581886.

\bibitem{Spruck:2020voo}
B.~Spruck \emph{et~al.}, ``{Belle II Pixel Detector Commissioning and
  Operational Experience},'' \emph{PoS}, vol. Vertex2019, p. 015, 2020.

\bibitem{Baehr:2015tbf}
S.~Baehr, O.~Sander, M.~Heck, C.~Pulvermacher, M.~Feindt, and J.~Becker,
  ``{Online-Analysis of Hits in the Belle-II Pixeldetector for Separation of
  Slow Pions from Background},'' \emph{J. Phys. Conf. Ser.}, vol. 664, no.~9,
  p. 092001, 2015.

\end{thebibliography}
